\documentclass{article}[12pt]
\usepackage{amsfonts,amsmath}

\newtheorem{theorem}{Theorem}

\def\R{{\mathbb R}}
\def\C{{\mathbb C}}

\begin{document}

\title{Two-dimensional von Neumann--Wigner potentials with
a multiple positive eigenvalue}

\author{R.G. Novikov \thanks{CNRS (UMR 7641), Centre de Math\'ematiques
Appliqu\'ees, Ecole Polytechnique, 91128 Palaiseau, France, and
 Faculty of Control and Applied Mathematics, MIPT, 141700 Dolgoprudny ,
Russia; e-mail:
novikov@cmap.polytechnique.fr} \and I.A. Taimanov
\thanks{Sobolev Institute of Mathematics, 630090 Novosibirsk,
Russia; e-mail: taimanov@math.nsc.ru} \and S.P. Tsarev
\thanks{Siberian Federal University,
Institute of Space and Information Technologies, 26 Kirenski
str., 660074 Krasnoyarsk, Russia; e-mail: sptsarev@mail.ru} }

\date{}

\maketitle

Let $H$ be a Schr\"odinger operator
$$
H = -\Delta + U
$$
with a potential $U(x)$, on $\R^N$, decaying at infinity. The
potential $U$ is called a {\it von Neumann--Wigner potential} if $H$
has a positive eigenvalue with an eigenfunction from $L_2(\R^N)$,
i.e. there is a point of discrete spectrum which is embedded into
the absolutely continuous spectrum.

The very first example of such a potential was constructed by von
Neumann and Wigner \cite{NW}. They found a three-dimensional
rotation-symmetric nonsingular potential $U(r)$ with the following
asymptotic behavior (a computational mistake in \cite{NW} reported
later is corrected here):
$$
U(r) = -\frac{8 \sin{2r}}{r} + O(r^{-2}) \ \ \ \ \mbox{as $r= |x|
\to \infty$, $ x \in \R^3$}.
$$
The Schr\"odinger operators with $U(x) = o(1/|x|)$ as $x \to \infty$
have no positive eigenvalues \cite{Kato}.

In the present note we construct multiparametric families of
explicit two-dimensional potentials which decay as $1/|x|$ and
have a multiple positive eigenvalue. To our knowledge these
are the first examples of such potentials.

We use the method introduced in \cite{TT0,TT} where two-dimensional
Schr\"odinger operators with fast decaying potentials and
multidimensional kernels were constructed. This method is based on
the Moutard transformation, of two-dimen\-sio\-nal Schr\"odinger
operators, which is as follows: let
$\omega$ be a formal solution of the equation
\begin{equation}
\label{1}
H\omega = \big(-\Delta + U(x,y)\big)\omega = 0, \quad
\Delta = \frac{\partial^2}{\partial x^2} +
\frac{\partial^2}{\partial y^2}.
\end{equation}
The Moutard transformation corresponding to $H$ and $\omega$
gives a new Schr\"odinger operator
$$
\widetilde{H} = -\Delta + \widetilde{U}, \ \ \ \ \widetilde{U} = U -
2 \Delta \log \omega
$$
such that if $\varphi$ satisfies $H\varphi=0$, then a function
$\theta$ determined modulo $\frac{\mathrm{const}}{\omega}$ by the
consistent system
\begin{equation}
\label{2} (\omega\theta)_x =
-\omega^2\left(\frac{\varphi}{\omega}\right)_y, \qquad
(\omega\theta)_y = \omega^2\left(\frac{\varphi}{\omega}\right)_x,
\end{equation}
satisfies
$$
\widetilde{H}\theta = 0.
$$
So there are maps
$$
U \to M_\omega(U) = U -2\Delta \log \omega, \ \ \ \ \varphi \to
S_\omega(\varphi) = \{ \theta + \frac{C}{\omega}, C \in \C\}.
$$
Let us consider an operator $H$ and a pair of solutions to
(\ref{1}): $\omega_1$ and $\omega_2$. For every $\theta_1 \in
S_{\omega_1}(\omega_2)$ there is a function (a result of double
iteration of the Moutard transformation)
\begin{equation}
\label{3}
\widehat{U} = M_{\theta_1} M_{\omega_1} (U) -  U = - 2\Delta \log
(\theta_1\omega_1).
\end{equation}
The result of this iteration depends on the choice of
$\theta_1 \in S_{\omega_1}(\omega_2)$, i.e., on the
integration constant $C$ in (\ref{2}). Moreover the functions
$$
\psi_1 = \frac{1}{\theta_1}, \ \ \ \psi_2 = \frac{\omega_2}{\omega_1
\theta_1}
$$
satisfy the equation
$$
(-\Delta + M_{\theta_1} M_{\omega_1} (U))\psi = 0.
$$

In contrast to \cite{TT0,TT} where such a double iteration was
applied to the case $U=0$, we apply it to the constant potential
$U=-k^2, k \in \R$. Therewith $\omega_1$ and $\omega_2$ have to
satisfy to the Helmholtz equation
$$
-\Delta \omega = k^2\omega.
$$
A large set of solutions to this equation is given by functions
of the form
\begin{equation}\label{z-zbar}
\mathrm{Re}\, \left[\frac{\partial^m}{\partial \lambda^m}
\exp\left(i\frac{k}{2}\left(\lambda z +
\frac{\bar{z}}{\lambda}\right)\right)\right], \quad z=x+iy, \ \
\lambda \in \C, \quad m =0,1,2,\dots,
\end{equation}
and their linear combinations.

For simplicity we consider the case
$k^2=1$ and demonstrate the method by one explicit example.

\begin{theorem}
Let $U=-1$ and
$$
\omega_1=x^2 \cos y -y \sin y  + y^2 \sin x +x \cos x ,  \ \ \
\omega_2=4(y\cos x +x\sin y), \ \ x,y \in \R.
$$
Then the two-dimensional potential $\widehat{U}$ takes the form
$$
\widehat{U} =  \frac{P}{Q^2}
$$
where
$$
Q = \omega_1 \theta_1 = -x^4-y^4 - 4 x^2 y \sin  x  \sin  y +x^2
\big(-8 \cos  y \sin x -2 \sin^2 y -1\big) + {}
$$
$$
{}+4 x y^2\cos  x  \cos  y -16 x y \cos  x  \sin
  y +2 x \cos  x  \big(-8 \cos  y -\sin  x\big)+{}
 $$
 $$
{} + y^2 \big(-8
 \cos  y  \sin  x +2 \sin ^2 x -3\big)+2 y \sin  y  \big(\cos  y
+8 \sin  x \big)+{}
$$
$$
{}+ 16 \cos  y  \sin  x +\sin ^2 x-\sin^2 y+4 \,C\,+1,
$$
and $P$ is a polynomial in $x,y$ and in sines and cosines of $x$ and
$y$:
$$
P = 16 \big(x^6 y \sin  x  \sin  y - x^5 y^2 \cos  x  \cos  y +{}
$$
$$
{}+x^2 y^5 \sin  x  \sin  y - x y^6 \cos  x  \cos  y \big) + (\dots)
$$
where by dots we denote lower order terms in $x$ and $y$. The
functions $\psi_1$ and $\psi_2$ take the form
$$
\psi_1 = \frac{\omega_1}{Q}, \ \ \ \ \psi_2 = \frac{\omega_2}{Q}
$$
and satisfy the equation
$$
\widehat{H}\psi = \psi \ \ \ \ \mbox{with $\widehat{H} = -\Delta +
\widehat{U}$.}
$$

Let us assume that $C$ is negative and $|C|$ is sufficiently large,
then $Q$ has no zeroes and therefore the potential $\widehat{U}$ and
the functions $\psi_1$ and $\psi_2$ are smooth. We have
\begin{equation}\label{as}
\widehat{U} = O\left(\frac{1}{r}\right), \ \ \ \psi_1 =
O\left(\frac{1}{r^2}\right), \ \ \ \psi_2 =
O\left(\frac{1}{r^3}\right), \ \ \ \mbox{as $r=\sqrt{x^2+y^2}
\to\infty$}.
\end{equation}
and therefore $\psi_1$ and $\psi_2$ lie in $L_2(\R^2)$ and are
linearly independent eigenfunctions of the operator $\widehat{H} =
-\Delta + \widehat{U}$ with the eigenvalue $E=1$.
\end{theorem}

Using various linear combinations of the solutions (\ref{z-zbar})
one can easily construct multiparametric families of similar
two-dimensional potentials $\widehat{U}$ with the asymptotics
(\ref{as}) and solutions $\psi_i$ at the energy level $E=k^2$ and even
improve the decay of the eigenfunctions $\psi_i$.
It is impossible to improve the decay of potentials
due to the already mentioned Kato theorem \cite{Kato}.

We guess that by applying multiple iterations one may obtain such
potentials with positive eigenvalues with higher multiplicity.

{\bf Acknowledgement.}
This work was done during the visit of one of the
authors (I.A.T.)  to Centre de Math\'ematique Appliqu\'ees of Ecole
Polytechnique in July, 2013.
The work was partially supported by TFP No 14.A18.21.0866 of
Ministry of Education and Sciences of
 Russian Federation  (R.G.N.),
the interdisciplinary
project ``Geometrical and algebraic methods of finding explicit
solutions to equations of mathematical physics and continuum mechanics''
of SB RAS (I.A.T. and S.P.T.) and by
the grant 1431/GF of Ministry of Education and Science of
Republic of Kazakhstan (I.A.T.).

\end{document}